\documentclass[aps,prb,twocolumn,floatfix,showpacs,superscriptaddress,fleqn]{revtex4-1}
\usepackage{amssymb}

\usepackage{graphicx}% Needed to include png/jpg/pdf figure files
\usepackage{amsmath}
\usepackage{bm}
\usepackage{color}

\usepackage[firstpage]{draftwatermark}%only first page
\usepackage[breaklinks]{hyperref}%hyperef package does internal and external linking (labels, urls)
\usepackage[all]{hypcap}%addition to hyperref, it ensures figure is correctly displayed (and not only the caption) when link is clicked

\hypersetup{
    plainpages=false,
    bookmarks=false,         % show bookmarks bar?
    unicode=false,          % non-Latin characters in Acrobat's bookmarks
    pdfborder={0 0 0}
    pdftoolbar=true,        % show Acrobat's toolbar?
    pdfmenubar=true,        % show Acrobat's menu?
    pdffitwindow=false,     % window fit to page when opened
    pdfstartview={FitH},    % fits the width of the page to the window
    pdftitle={LNCMI Annual Report 2012},    % title
    pdfauthor={Duncan Maude},     % author
    pdfproducer={LNCMI-CNRS-UJF-UPS-INSA}, % producer of the document
    pdfkeywords={High} {Magnetic} {Fields}, % list of keywords
    pdfnewwindow=true,      % links in new window
    linktoc=section,
    colorlinks=true,       % false: boxed links; true: colored links
    linkcolor=blue,          % color of internal links
    citecolor=red,        % color of links to bibliography
    filecolor=magenta,      % color of file links
    urlcolor=blue           % color of external links
}

\begin{document}
%\affiliation{Laboratoire National des Champs Magn\'etiques Intenses, UPR 3228, CNRS-UJF-UPS-INSA, Grenoble and
%Toulouse, France}

\title{Second order resonant Raman scattering in single layer tungsten disulfide (WS$_{2}$)}

\author{A. A. Mitioglu}
\affiliation{LNCMI, CNRS-UJF-UPS-INSA, Grenoble and Toulouse, France} \affiliation{Institute of Applied Physics,
Academiei Str. 5, Chisinau, MD-2028, Republic of Moldova}

\author{P. Plochocka}
\affiliation{LNCMI, CNRS-UJF-UPS-INSA, Grenoble and Toulouse, France}\email{paulina.plochocka@lncmi.cnrs.fr}

\author{G. Deligeorgis }

\affiliation{FORTH-IESL, Microelectronics Research Group, P.O. Box 1527, 71110 Heraklion, Crete, Greece}

\author{S. Anghel}
\affiliation{Institute of Applied Physics, Academiei Str. 5, Chisinau, MD-2028, Republic of Moldova}
\affiliation{Ruhr-Universit\"at Bochum, Anorganische Chemie III, D-44801 Bochum Germany}

\author{L. Kulyuk}
\affiliation{Institute of Applied Physics, Academiei Str. 5, Chisinau, MD-2028, Republic of Moldova}

\author{D.~K.~Maude}
\affiliation{LNCMI, CNRS-UJF-UPS-INSA, Grenoble and Toulouse, France}

\SetWatermarkText{FIRST DRAFT PLEASE DO NOT DISTRIBUTE}%Text for watermark
\SetWatermarkLightness{0.9}%How light/dark
\SetWatermarkScale{0.25}%Font scaling

\date{\today}

\begin{abstract}
Resonant Raman spectra of single layer WS$_{2}$ flakes are presented. A second order Raman peak (2LA) appears under
resonant excitation with a separation from the E$^{1}_{2g}$ mode of only $4$cm$^{-1}$. Depending on the intensity ratio
and the respective line widths of these two peaks, any analysis which neglects the presence of the 2LA mode can lead to
an inaccurate estimation of the position of the E$^{1}_{2g}$ mode, leading to a potentially incorrect assignment for
the number of layers. Our results show that the intensity of the 2LA mode strongly depends on the angle between the
linear polarization of the excitation and detection, a parameter which is neglected in many Raman studies.
\end{abstract}

\maketitle

%\SetWatermarkText{FIRST DRAFT PLEASE DO NOT DISTRIBUTE}%Text for watermark
%\SetWatermarkLightness{0.9}%How light/dark
%\SetWatermarkScale{0.25}%Font scaling

%Keywords: Single layer WS$_2$;  Resonant Raman spectroscopy;

\section{Introduction}

Single layer transition metal dichalcogenides (TMDs) are a class of emerging nano-materials which have recently become
the subject of intense investigation, in part due to their truly two-dimensional character and due to the wealth of
applications that may benefit from their unique characteristics. In TMDs, a single layer is composed of a monolayer of
the transition metal with a chalcogen monolayer above and below. The intra layer coupling is strong due to the
chalcogen metal covalent bonds. The electronic properties of a single layer are significantly different from bulk
crystals, where the stacked layers are coupled by weak Van deer Waals forces. Bulk TMDs are semiconductors with an
indirect gap in the near infrared spectral range. In contrast, single layer TMDs, such as molybdenum disulfide
(MoS$_{2}$), tungsten disulfide (WS$_{2}$) or tungsten diselenide (WSe$_{2}$), are two dimensional (2D) semiconductors
with \emph{a direct gap in the visible spectral range} with a large number of potential applications in
optoelectronics.~\cite{Mak10,Splendiani10,Eda11,Albe02,Gutierrez13,Zhao13,Wang12,Cao12,Mak13}

Many experimental techniques have been employed to study the electronic properties of layered materials. Among them,
Raman spectroscopy, has been extensively used to determine the number of layers in
TMDs~\cite{Gutierrez13,Li12,Lee10,Li13,Wang12} and in graphene.~\cite{Ferrari06,Gupta06} The vibrational spectra are
sensitive to the thickness of the crystal so that small changes in the separation of first order Raman modes can be
used to differentiate between bulk, monolayer, bilayer and trilayer crystals. In transition metal dichalcogenides, the
excitation of Raman modes can easily be tuned to coincide with one of the direct optical transitions. Indeed, the large
number of transitions (excitons A,B and C), which are often fortuitously close to laser lines commonly used, means that
Raman data is often taken close to resonant conditions. In turn this can result in additional and frequently unwanted
complications when interpreting the data.

Resonant Raman spectra are composed of both first and second order Raman excitations, as demonstrated for bulk
MoS$_{2}$~\cite{Chen74,Frey99} or WS$_{2}$.~\cite{Sourisseau89,Sourisseau91,Sekine80} Recently, second order Raman
modes have been reported in single layer WS$_{2}$,~\cite{Berkdemir13,Dumenco11,Song13,Zeng13} MoS$_{2}$~\cite{Li12a},
TaSe$_{2}$~\cite{Hajiyev13} and WSe$_{2}$.~\cite{Luo13} Such resonant vibrational spectra can provide significant
additional information concerning the electronic properties. For example, it has been suggested that the observation of
a strong second order Raman resonance involving the longitudinal acoustic phonons (2LA) in monolayer WS$_{2}$ is a
signature of the single layer nature of the sample.~\cite{Berkdemir13}

Nevertheless, the resonant Raman spectra in atomically thin TMDs are relatively unexplored and far from being fully
understood. In monolayer WSe$_{2}$, an additional Raman mode, separated by only a few cm$^{-1}$ from the E$^{1}_{2g}$
and A$_{1g}$ modes, and whose origin remains to be elucidated, was recently reported.~\cite{Luo13}
%Generally speaking, using resonant excitation can lead to a potentially incorrect determination of the thickness of the
%crystal if the presence of second order Raman modes leads to errors when determining the \textbf{exact} separation of
%the first order Raman modes. This is especially problematic  if the second order vibrational peak is very close to a
%first order peak, as for example in WS$_{2}$, WSe$_{2}$, TaSe$_{2}$.
From a fundamental physics point of view, the inter valley scattering process involving the 2LA phonon was suggested to
be the main source of valley depolarization.\cite{Kioseoglou12} For all these reasons, understanding the resonant Raman
processes in TDMs is of great importance.

In this paper, we present resonant Raman scattering measurements performed on a single layer of tungsten disulphide. We
focus on the second order Raman resonance involving the longitudinal acoustic phonon (2LA), which is separated by only
$4$cm$^{-1}$ from the E$^{1}_{2g}$ mode. We show that due to their small separation, the two modes can be
undistinguishable, leading to an erroneous estimation of the separation of the E$^{1}_{2g}$ and A$_{1g}$ modes, which
is normally accepted to be a robust indication of the crystal thickness. Crucially, measurements involving linear
polarization show that depending on the angle between polarization of the excitation and detection the 2LA mode can
even dominate over the E$^{1}_{2g}$ mode or vice versa. In most Raman measurements performed to characterize samples,
the polarization of the excitation and detection is unknown so that incorrect conclusions concerning the number of the
layers present can easily be reached.

\section{Experimental Techniques}

Single bulk crystal of 2H-WS$_2$ (the hexagonal 2H-polytype of tungsten disulphide) have been grown using chemical
vapor transport with Bromine as the transport agent.~\cite{Mitioglu13} Single layer flakes of tungsten disulfide have
been obtained from these crystals using mechanical exfoliation. For the measurements, the sample was placed in a helium
flow cryostat with optical access. The $\mu$-Raman measurements have been performed in the back scattering
configuration. Excitation and collection was implemented using a microscope objective with a numerical aperture $NA =
0.66$ and magnification $50\times$. The typical diameter of the laser spot was approximately $1 \mu$m. The $\mu$-Raman
spectra have been recorded using a spectrometer equipped with a CCD camera and a green solid-state laser, emitting at
$532$ nm, was used for excitation.

\section{Resonant Raman Scattering}

The 2H-WS$_2$ crystal belongs to $D^{4}_{6h}$ space group. Of the $12$ first order phonon modes at the Brillouin zone
center four are Raman active: $A_{1g}$, E$_{1g}$, E$^{1}_{2g}$ and E$^{2}_{2g}$. In the back scattering configuration,
the E$_{1g}$ mode is forbidden, and the E$^{2}_{2g}$ mode is observed only in the very low frequency
region.~\cite{Sekine80} In addition, E$^{2}_{2g}$ is a shared mode originating from the relative motion of the atoms in
different layers, and is thus absent in single layer samples. As a result, the Raman spectra of a single layer WS$_2$
are dominated by two first order modes: E$^{1}_{2g}$, which is associated with the in plane motion of the sulfur and
tungsten atoms in opposite directions and $A_{1g}$, which is associated with the out of plane motion of the sulfur
atoms.~\cite{Sekine80} In bulk WS$_2$, the separation between these two modes is $66$cm$^{-1}$ and it decreases
gradually with the number of layers reaching $62$cm$^{-1}$ for a single
layer.~\cite{Gutierrez13,Berkdemir13,Zhao13,Zeng13,Song13}

\subsection{Room temperature}

\begin{figure}
\begin{center}
\includegraphics[width= 8.5cm]{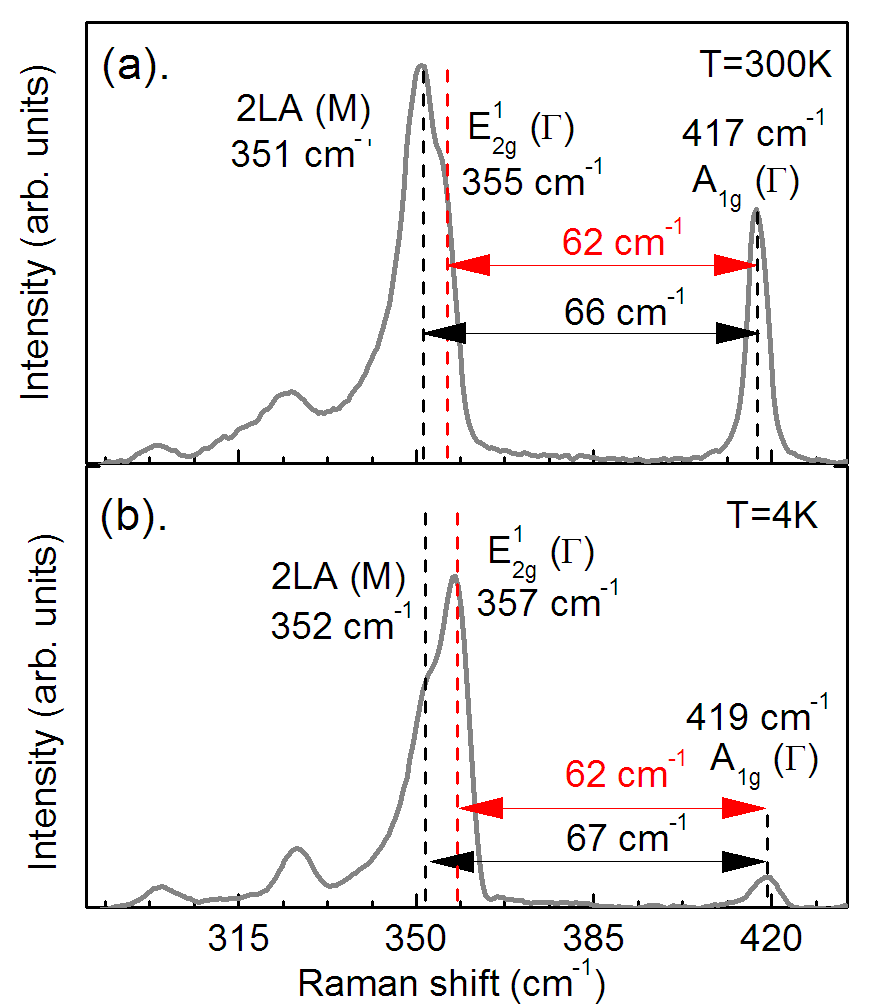}
\end{center}
\caption{(color online) (a), (b) Typical $\mu$Raman spectra measured at $T=300$ and $T=4$K. The dashed lines indicate
the position of three Raman modes: 2LA, E$^{1}_{2g}$ and A$_{1g}$.}\label{Fig1}
\end{figure}

A typical resonant Raman spectrum taken at 300K is presented in Fig~\ref{Fig1}(a). Two strong features are observed. A
single resonance peak at 417 cm$^{-1}$, which corresponds to the $A_{1g}$ mode in single layer of tungsten
disulphide,~\cite{Gutierrez13,Berkdemir13,Zhao13,Zeng13,Song13} together with a more complicated feature observed
around 353 cm$^{-1}$. It is composed of two features; a peak at 351 cm$^{-1}$ accompanied by a weaker peak at 355
cm$^{-1}$ corresponding to the E$^{1}_{2g}$ mode in single layer
WS$_{2}$.~\cite{Gutierrez13,Berkdemir13,Zhao13,Zeng13,Song13} The separation between $A_{1g}$ and E$^{1}_{2g}$ of
62cm$^{-1}$ proves the single layer character of our sample. The presence of an additional peak on the low frequency
side of the E$^{1}_{2g}$ mode has been reported for bulk crystals~\cite{Sourisseau89,Sourisseau91,Sekine80} and very
recently for single layer WS$_{2}$.~\cite{Berkdemir13,Song13} This additional peak has been assigned to a second order
Raman resonance involving longitudinal acoustic phonons
(2LA).~\cite{Sourisseau89,Sourisseau91,Sekine80,Berkdemir13,Song13} The peak we observe at 351 cm$^{-1}$ corresponds
well to such a second order Raman resonance. Our data suggests, in apparent agreement with calculations and
measurements presented by Berkdemir et al.,~\cite{Berkdemir13} that the 2LA mode is stronger than E$^{1}_{2g}$ mode for
single layer material. We will see later that this agreement is fortuitous since the angle of the linear polarization
between excitation and detection is completely unknown in our measurements.

\subsection{Low temperature}

Further proof of the 2LA character of the peak at 351 cm$^{-1}$ is provided by measurements at low temperature ($T=4$K)
presented in Fig~\ref{Fig1}(b). Compared to the Raman data measured at 300K, a small blue shift of the resonance
frequencies is observed with decreasing temperature due to the anharmonic vibrations in the lattice in the interatomic
potential energy mediated by phonon - phonon interactions~\cite{Rusen14,Lanzillo13}.
%A similar behavior, reported for single layer MoS$_{2}$, was attributed to the anharmonic vibrations in the lattice in
%the interatomic potential energy mediated by phonon - phonon interactions~\cite{Rusen14,Lanzillo13}. The observed
%decrease of the intensity of the A$_{1g}$ mode with temperature is also in agreement with a previous investigation of
%bulk WS$_{2}$~\cite{Sourisseau91}.
However, the dominant effect of the temperature is the radical change in the intensity ratio between the E$^{1}_{2g}$
and 2LA modes. At room temperature the 2LA mode dominates over the E$^{1}_{2g}$ while at low temperature the situation
is reversed. As the 2LA mode is an overtone it's temperature dependence is expected to take the form
$[n(\omega_{1},T)+1]\times[n(\omega_{2},T)+1]$, where $n(\omega,T)=(\exp(\frac{\hbar\omega}{kT})-1)^{-1}$ is the phonon
occupation number for a process in which phonons with frequencies $\omega_{1}, \omega_{2}$ are absorbed and created
respectively.~\cite{Chen74} A quick estimation using the frequency of 2LA mode shows that it's intensity should
decrease by 1.5 times between room temperature and 4K. This estimation is in a good agreement with our data with the
assumption that the intensity of the E$^{1}_{2g}$ mode is rather insensitive to temperature.~\cite{Sourisseau91} Our
data is clearly not consistent with the assignment of the 351 cm$^{-1}$ feature to a combination process whose
temperature dependence would take the form $n(\omega_{1},T)\times[n(\omega_{2},T)+1]$.~\cite{Chen74} In such a scenario
the resonance would simply vanish at 4K. The temperature dependence of the intensity of this resonance is therefore an
independent proof of it's 2LA character. We note that the variation of the direct gap with temperature could detune the
resonant excitation and also in principle contribute to a decrease of the intensity of the 2LA resonance. However, the
phonon occupation is expected to dominate the temperature dependence.

\subsection{Assigning layer thickness}

\begin{figure}[]
\begin{center}
\includegraphics[width= 8.5cm]{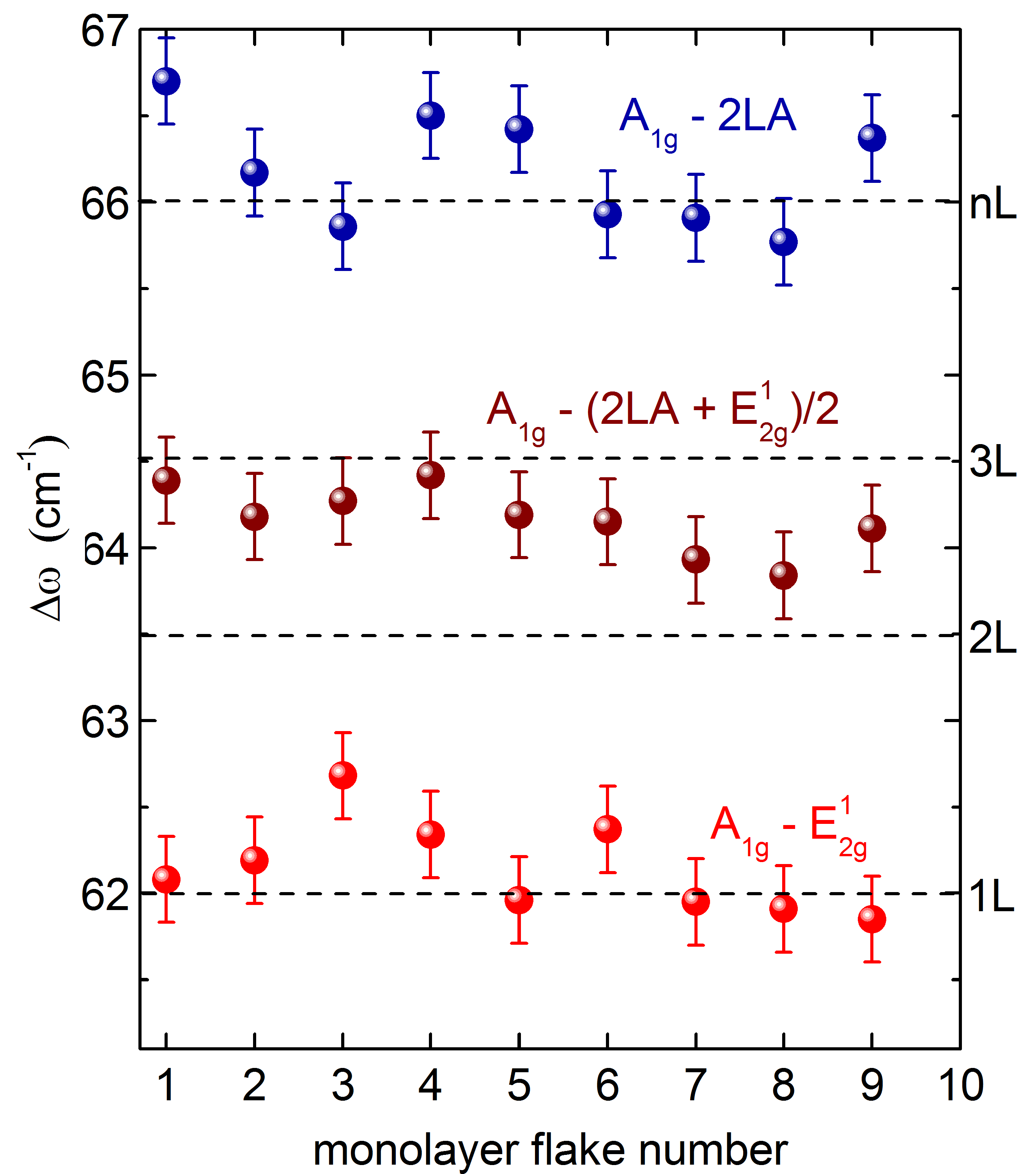}
\end{center}
\caption{(color online) The symbols indicate the frequency difference $\Delta\omega$ between: $A_{1g}$ and $2LA$,
between $A_{1g}$ and the average frequency of $2LA$ and $E^{1}_{2g}$, and between $A_{1g}$ and $E^{1}_{2g}$. The dashed
lines indicate the accepted values for the separation of $A_{1g}$ and $E^{1}_{2g}$ for single, bi -, tri-layer and
bulk.~\cite{Molina11,Berkdemir13,Zhao13}}\label{Fig2}
\end{figure}

Let us now focus on the separation between these three resonances. Regardless of the temperature, the separation
between A$_{1g}$ and E$^{1}_{2g}$ is 62 cm$^{-1}$ and the separation between A$_{1g}$ and 2LA is 66 cm$^{-1}$. However,
the separation between the 2LA and E$^{1}_{2g}$ modes is only $4$cm$^{-1}$ so that, any broadening of the peaks due to
impurities/defects in the sample or simply a lack of resolution might lead to an incorrect determination of the
frequency of the E$^{1}_{2g}$ mode. As a result, the separation between A$_{1g}$ and E$^{1}_{2g}$  might be
overestimated, in turn leading to an overestimation of the  number of layers in the sample. To illustrate the
universality of this problem, we have measured Raman spectra for several single layer flakes in order to analyze the
separation $\Delta\omega$ between different Raman features. The summary of such an analysis is presented in
Fig~\ref{Fig2}. The symbols correspond to the three different combinations: $\Delta\omega_{1} = \omega(A_{1g}) -
\omega(2LA)$, $\Delta\omega_{2} = \omega(A_{1g}) - (\omega(2LA)+\omega(E^{1}_{2g}))/2$ and $\Delta\omega_{3} =
\omega(A_{1g})- \omega(E^{1}_{2g})$ respectively. The latter is generally used in the literature to determine the
number of layers present. For comparison, the literature values for the separation between (A$_{1g}$ and E$^{1}_{2g}$
for single, bi -, tri-layer and bulk\cite{Molina11,Berkdemir13,Zhao13} are indicated by broken lines.

The Raman splitting $\Delta\omega_{3}$ is generally considered to be a robust signature of the number of the layers in
the sample. For our data, in which E$^{1}_{2g}$ and 2LA are clearly resolved, for all the flakes, $\Delta\omega_{3}$ is
equal to $62$cm$^{-1}$, demonstrating the single layer character of our flakes.~\cite{Molina11,Berkdemir13,Zhao13}
However, if the modes E$^{1}_{2g}$ and 2LA are not well resolved, in any analysis which neglects the possible presence
of the 2LA mode the assignment of the number of layers becomes highly problematic. If the 2LA mode dominates over
E$^{1}_{2g}$ the measured difference will correspond to the $\Delta\omega_{1} = \omega(A_{1g}) - \omega(2LA)$ which is
approximately $66$cm$^{-1}$, leading to the incorrect assignment of Raman from a bulk
sample.~\cite{Sourisseau89,Sourisseau91,Sekine80} On the other hand, if the intensity of 2LA and E$^{1}_{2g}$ modes are
similar, the frequency of the Raman features will correspond to the average of the two peaks. The measured frequency
difference $ \omega(A_{1g}) - (\omega(2LA)+\omega(E^{1}_{2g}))/2 \simeq 64$cm$^{-1}$ then correspond to the bi- or
tri-layer system.~\cite{Molina11,Berkdemir13,Zhao13} This poses a significant problem for the correct assignment of the
number of layers especially since the relative intensity between between E$^{1}_{2g}$ and 2LA depends on several
parameters, including the temperature and the excitation wavelength (2LA is only observed for resonant excitation) as
shown for bulk crystals.\cite{Sourisseau91,Sourisseau89} In the following, we show that the relative linear
polarization of the excitation and detection also significantly influences this intensity ratio.

\begin{figure*}[]
\begin{center}
\includegraphics[width= 14 cm]{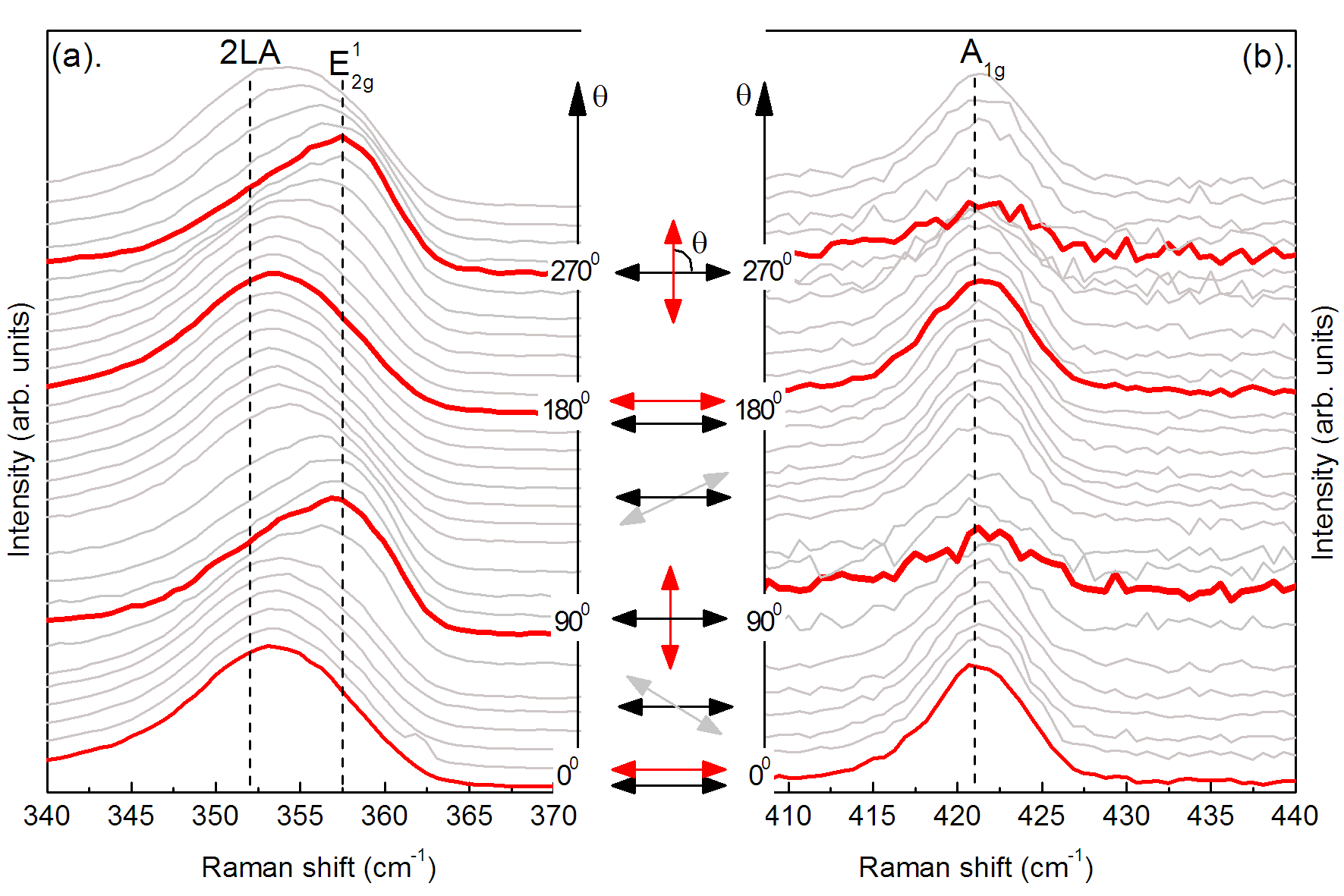}
\end{center}
\caption{(color online) Typical results are presented in (a) for the frequency region corresponding to the 2LA and
E$^{1}_{2g}$ modes and in (b) for the A$_{1g}$ mode. The spectra for different angles $\theta$ are vertically shifted
for clarity. The spectra for co and cross polarized configuration are indicated by thick red lines. The orientation of
the detection (red arrows) and excitation (black arrows) linear polarization are indicated for the co and cross
polarization configuration.}\label{Fig3}
\end{figure*}

\subsection{Polarization dependence}

$\mu$-Raman spectra have been measured at $T=4.2$K as a function of the relative angle $\theta$ between the linearly
polarized excitation and the emission. We have performed measurements when the linear polarization of the excitation
was fixed and the angle was varied in the detection and vice versa. We have verified that the results are only
dependent on the angle $\theta$; the same spectra are always observed for co and cross linearly polarized beams
independently of the angle of the linear polarization with respect to the sample. Typical results are presented in
Fig~\ref{Fig3}(a) for frequencies in the region corresponding to the 2LA and E$^{1}_{2g}$ modes and in
Fig~\ref{Fig3}(b) for the A$_{1g}$ mode. The spectra for different angles are shifted vertically for clarity. The
spectra for co and cross polarized configuration are indicated by thick red lines. The intensity of the 2LA mode
changes as a function of the angle $\theta$. When the polarization of the excitation and detection are orthogonal, the
intensity of 2LA phonon resonance reaches a minimum.  For parallel configuration the 2LA mode is stronger than
E$^{1}_{2g}$. For angles $0<\theta < 90$ degrees the intensity of 2LA mode changes gradually. A similar behavior is
observed for A$_{1g}$ mode while the intensity of E$^{1}_{2g}$ is constant within experimental error.

\begin{figure}[]
\begin{center}
\includegraphics[width= 8.5cm]{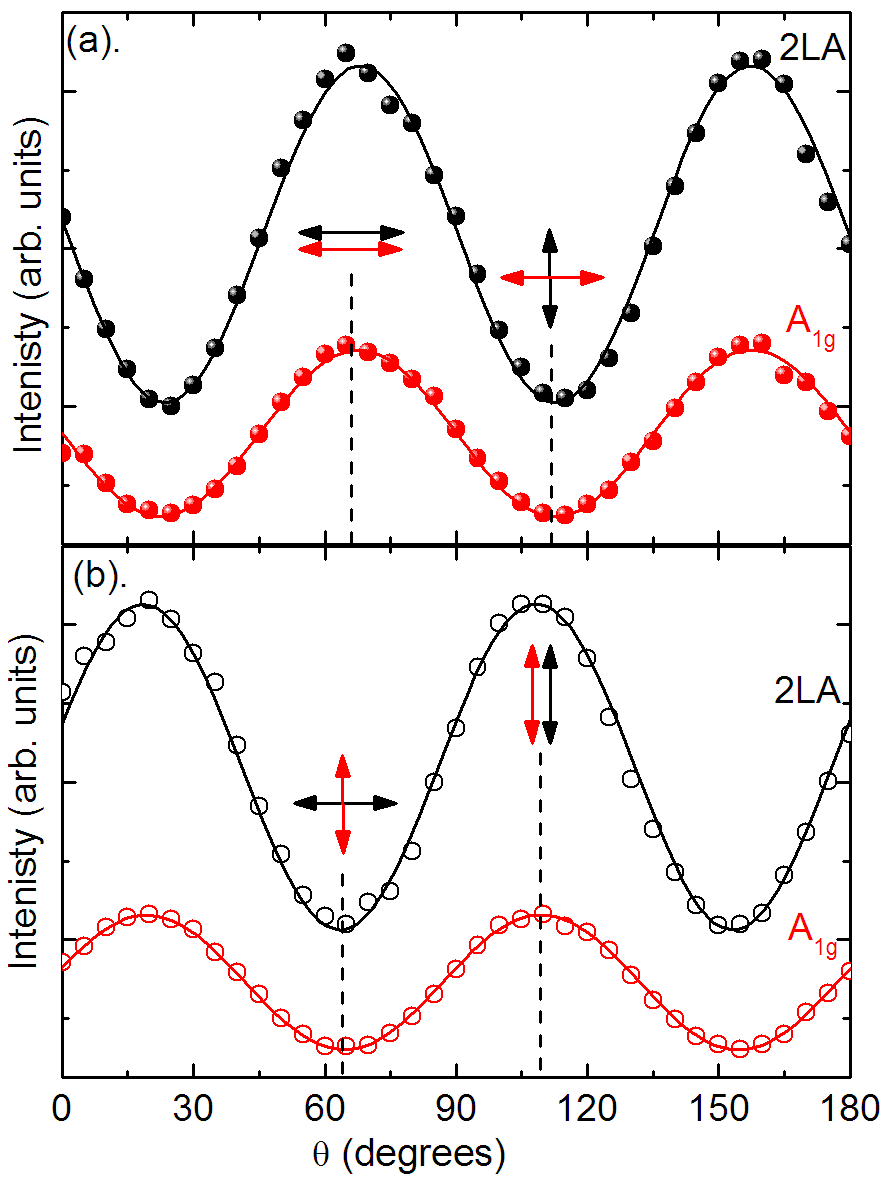}
\end{center}
\caption{(color online) Intensity of the 2LA and  A$_{1g}$ modes (symbols). (a), (b) represent two different
experimental configurations; the linear polarization in detection was either horizontal or vertical as indicated by the
red arrows. The black arrows indicated the excitation polarization in the co and crossed configurations }\label{Fig4}
\end{figure}

The intensities of the Raman peaks as a function of the angle between polarization of the excitation and detection are
presented in Fig~\ref{Fig4}.
%To verify that the only important parameter is the angle between linearly polarized light in excitation and detection,
%rather than the angle of polarization with respect to the sample, we have performed two distinct series of
%measurements.
The linear polarization in detection was fixed, either horizontal or vertical as indicated by the red arrows on
Fig~\ref{Fig4}(a)-(b). The polarization of the excitation was then rotated through $180^{0}$. The co and cross
polarized configurations are indicated by the arrows in Fig~\ref{Fig4}(a)-(b). The black arrow represent the
orientation of the polarization of the excitation. The data obtained for horizontal and vertical configurations are
presented in Fig~\ref{Fig4}(a)-(b) respectively. The experimental values of the intensity of 2LA and A$_{1g}$ modes are
indicated by symbols. The data is well described by a sinusoidal function (solid line). The minimum of the intensity of
both modes for horizontally polarized detection is shifted by 90 degrees with respect to the results for the vertically
polarized detection. This proves that the ratio between 2LA and E$^{1}_{2g}$ depends only on the angle $\theta$ between
the excitation and detection polarization. For the A$_{1g}$ mode the oscillation of the intensity as a function of the
angle $\theta$, as well as the lack of intensity changes of E$^{1}_{2g}$ (not shown) is understood and comes from the
form of Raman tensors.~\cite{Loudon64} The analysis for the 2LA mode is not as straight forward as that of the
A$_{1g}$, since the 2LA mode arises from a double resonance process. However, it is experimentally well established
that the A$_{1g}$ and 2LA modes have the same polarization dependence. For example, the polarization characteristic of
2LA mode has been determined in bulk crystals \cite{Sekine80,Sourisseau91,Sourisseau89} where it was compared with the
results of neutron scattering where the phonon dispersion along crystal axis was measured in the same plane as the
polarization of the light. As already demonstrated by Chen \emph{et al.}~\cite{Chen14} a single layer, exhibits the
same polarization characteristic.

\subsection{Raman modes at lower energies}

Finally, we will discuss the two Raman modes with low intensity observed at $\sim 300$ and $\sim 320$ cm$^{-1}$ in
Fig.\ref{Fig1}(a-b). They have been contradictorily linked to the monolayer or multilayer nature of the flake
studied.~\cite{Zhao2013a,Berkdemir13} We have investigated a large number of samples and these modes are present in all
investigated flakes which show single layer character. They could be assigned to 2LA modes with a wave vector in the
vicinity of the K point in the Brillouin zone as suggested for bulk crystals.\cite{Sourisseau89} An alternative
assignment would be the E$_{1g}$ mode, which in bulk crystals is not allowed in the back scattering (reflection)
geometry. Certainly, this controversy shows that resonant Raman spectroscopy is far from being fully understood in
single layer dichalcogenides.

\section{Conclusion}

To summarize, under resonant excitation, the presence of the 2LA Raman mode complicates the determination of the number
of layers in WS$_2$ flakes. Thankfully, the assignment of a monolayer from the E$^{1}_{2g}$-A$_{1g}$ separation remains
robust. However, mono layers can be incorrectly assigned to bi-layer, tri-layer or even bulk crystals depending on the
intensity of the 2LA mode. All our data has been taken using 532nm excitation, nevertheless, our conclusions should
apply to other excitation wavelengths. Sourisseau \emph{et al.}~\cite{Sourisseau89,Sourisseau91} studied in detail the
dependence of the intensity of 2LA mode as a function of excitation wavelength. They have shown, that whenever the
laser is in resonance with the excitonic transition (A,B or C) both the 2LA and the A$_{1g}$ modes are enhanced. Thus,
our conclusions and analysis will be valid whenever the 2LA mode is present in the Raman spectrum, in other words
whenever the laser is in resonance with one of the excitonic transitions. The evolution of the intensity ratio of the
2LA and E$^{1}_{2g}$-A$_{1g}$ modes with polarization prescribes using the intensity of the 2LA mode to identify a
monolayer, unless the polarization angle between excitation and detection is carefully controlled, which is rarely the
case. As many Raman characterizations are frequently performed close to resonant excitation, and without controlling
the polarization angle, considerable care should be exercised when analyzing these spectra.

\begin{acknowledgments}
This work was partially supported by  Programme Investissements d'Avenir under the program ANR-11-IDEX-0002-02 -
reference ANR-10-LABX-0037-NEXT, ANR JCJC project milliPICS, the Region Midi-Pyr\'en\'ees under contract MESR 13053031
and STCU project 5809.
\end{acknowledgments}

%\bibliography{Ws2bib}

\end{document}